\documentclass{svmult}

\usepackage{makeidx}         
\usepackage{graphicx}        
\usepackage{multicol}        

\makeindex             

\begin{document}

\title*{Five Years of Continuous-time Random Walks in Econophysics}
\author{Enrico Scalas}
\institute{DISTA, Universit\`a del Piemonte Orientale, Via Bellini 25/G, 15100 Alessandria, Italy and INFM Unit\`a di Genova, Via Dodecaneso 33, 16146 Genova. Italy
\texttt{scalas@unipmn.it}}
%
%
\maketitle

\section*{Abstract}

This paper is a short review on the application of continuos-time random walks
to Econophysics in the last five years.

\section{Introduction}
\label{sec:1}





%
%
%
%
%
%
%

Recently, there has been an increasing interest 
on the statistical properties of high-frequency financial data related to market microstructural properties \cite{goodhart97,ohara99,madhavan00,dacorogna01,raberto01,luckock03}.
High-frequency econometrics is now well established after research on
autoregressive conditional duration models \cite{engle97,engle98,bauwens00,lo02}.

In high-frequency financial data not only returns but also waiting times between consecutive trades are random variables \cite{zumbach98}. This remark is present in a paper by Lo and McKinlay \cite{lo90}, but it can be traced at least to papers on the application of compound Poisson processes \cite{press67} and subordinated stochastic processes \cite{Clark 73} to finance. Models of tick-by-tick financial data based on compound Poisson processes can also be found in the following references: \cite{rydberg98,rydberg99,rydberg00}.

Compound Poisson processes are an instance of continuous-time random walks (CTRWs) \cite{montroll65}. The application of CTRW to economical problems dates back, at least, to the 1980s. In 1984, Rudolf Hilfer discussed the application of  stochastic processes to operational planning, and used CTRWs as tools for sale forecasts \cite{hilfer84}. The revisited and augmented CTRW formalism has been applied to high-frequency price dynamics in financial markets by our research group since 2000, in a series of three papers \cite{scalas00,mainardi00,gorenflo01}. Other scholars have recently used this formalism \cite{masoliver03a,masoliver03b,kutner03}. However, already in 1903, the PhD thesis of Filip Lundberg presented a model for ruin theory of insurance companies, which was later developed by Cram\'er \cite{lundberg03,cramer30}. The underlying stochastic process of the Lundberg-Cram\'er model is another example of compound Poisson process and thus also of CTRW.

Among other issues, we have studied the independence between log-returns and waiting times for stocks traded at the New York Stock Exchange in October 1999. For instance, according to a contingency-table analysis performed on General Electric (GE) prices, the null hypothesis of independence can be rejected with a significance level of 1 \% \cite{raberto02}. We have also discussed the anomalous non-exponential behaviour of the unconditional waiting-time distribution between tick-by-tick trades both for future markets \cite{mainardi00} and for stock markets \cite{raberto02,scalas04b}. Different waiting-time scales have been investigated in different markets by various authors. All these empirical analyses corroborate the waiting-time anomalous behaviour. A study on the waiting times in a contemporary FOREX exchange and in the XIXth century Irish stock market was presented by Sabatelli {\it et al.} \cite{sabatelli02}. They were able to fit the Irish data by means of a Mittag-Leffler function as we did before in a paper on the waiting-time marginal distribution in the German-bund future market \cite{mainardi00}. Kyungsik Kim and Seong-Min Yoon studied the tick dynamical behavior of the bond futures in Korean Futures Exchange (KOFEX) market and found that the survival probability displays a stretched-exponential form \cite{kim03}. Finally, Ivanov {\it et al.} \cite{ivanov04} confirmed that a stretched exponential fits well the survival distribution for NYSE stocks as we suggested in \cite{raberto02}. Moreover, just to stress the relevance of non-exponential waiting times, a power-law distribution has been recently detected by T. Kaizoji and M. Kaizoji in analyzing the calm time interval of price changes in the Japanese market \cite{kaizoji03}. We have offered a possible explanation of the anomalous waiting-time behaviour in terms of daily variable activity \cite{scalas04b}.

The aforementioned empirical results are important as market microstructural models should be able to reproduce such a non-exponential behaviour of waiting-time distributions in order to be realistic. However, the rest of this paper focuses on the theory and is divided as follows: in Section 2, CTRW theory is presented as applied to finance. Finally, in Sec. 3, a summary of main results is presented together with a discussion on the direction of future research.

\section{Theory}

Random walks have been used in finance since the seminal
thesis of Bachelier \cite{Bachelier 00}, a work completed at the end of the 
XIXth century, more than a hundred years ago. After a rather long 
period in which the ideas of Bachelier were neglected, they were further 
developed until recent times \cite{Cootner 64,Merton 90}.

Our approach to random walks in finance is related to 
that of Clark \cite{Clark 73} and to the
introductory part of Parkinson's paper \cite{parkinson77}. 
It is a purely phenomenological approach.
There is no assumption on the rationality or
the behaviour of trading agents and it is not necessary to assume the
validity of the efficient market hypothesis \cite{fama70,fama91}.
However, as briefly discussed above, even in the absence of a {\it microfoundation}, 
a phenomenological model can
still be useful to corroborate or falsify the consequences
of behavioural or other assumptions on markets. 
The model itself can be corroborated or falsified by empirical data.

In order to model tick-by-tick data, we use the so-called continuous-time random walk (CTRW), where time intervals 
between successive steps are random variables, as discussed by Montroll and 
Weiss \cite{montroll65}. In physics, CTRWs have been introduced as models
of diffusion with instantaneous jumps from one position to the next.
For this reason they can be used as models of price dynamics as well.

Let $S(t)$ denote the price of an asset at time
$t$. In a real market with a double-auction mechanism, 
prices are fixed when buy orders are matched with
sell orders and a transaction (trade) occurs. It is
more convenient to refer to returns rather than 
prices. For this reason, we shall take into account 
the variable $x(t) = \log S(t)$: the logarithm of the price.
For a small price variation $\Delta S = S(t_{i+1}) - S(t_{i})$, the 
return $r = \Delta S/S(t_{i})$ and the logarithmic return 
$r_{log} = log[S(t_{i+1}) / S(t_{i})]$ virtually coincide. 

CTRWs are essentially point processes with reward \cite{cox79}. The point process is characterized by  a sequence of independent identically distributed (i.i.d.) positive random variables $\tau_i$, which can be interpreted as waiting times between two consecutive events:
\begin{equation}
\label{timewalk}
t_n = t_0 + \sum_{i=1}^{n} \tau_i; \; \; t_n - t_{n-1} = \tau_n; \; \; n=1, 2, 3, \ldots;
\; \; t_0 = 0. \end{equation} 
The rewards are (i.i.d.) not necessarily positive random variables: $\xi_i$. In the usual physical intepretation, the $\xi_i$s represent the jumps of a diffusing particle, and they can be $n$-dimensional vectors. Here, only the 1-dimensional case is studied, but the extension of many results to the $n$-dimensional case is straightforward. The position $x$ of the particle at time $t$ is given by the following random sum (with $N(t) = \max \{ n:\;t_{n} \leq t \}$ and $x(0)=0$):
\begin{equation}
\label{jumpwalk}
x(t) = \sum_{i=1}^{N(t)} \xi_i.
\end{equation}
In the financial interpretation outlined above, the $\xi_i$'s have the meaning of log-returns, whereas the {\it positions} or rewards $x(t)$ represent log-prices at time $t$.
Indeed, the time series $\{ x(t_i) \}$
is characterised by $\varphi(\xi, \tau)$, the 
{\em joint probability density}
of log-returns $\xi_{i} = x(t_{i+1}) - x(t_{i})$ and of waiting times 
$\tau_i = t_{i+1} - t_{i}$. The joint density
satisfies the normalization condition 
$\int \int d \xi d \tau \varphi (\xi, \tau) = 1$. It must be again remarked that both $\xi_i$ and $\tau_i$ are assumed to be independent and identically distributed (i.i.d.) random variables. This strong assumption is useful to derive limit theorems for the stochastic processes described by CTRWs. However, in financial time series, the presence of volatility clustering, as well as correlations between waiting times do falsify the i.i.d hypothesis. The reader interested in a review on correlated random variables in finance is referred to chapter 8 in McCauley's recent book \cite{mccauley04}.

In general, log-returns and waiting times are not independent from each other \cite{raberto02}. By probabilistic arguments (see \cite{montroll65,mainardi00,scalas04}), one can derive the following integral equation that gives the probability density, $p(x,t)$, for the particle of being in position $x$ at time $t$, conditioned by the fact that it was in position $x=0$ at time $t=0$:
\begin{equation}
\label{masterequation}
p(x,t) =  \delta (x)\, \Psi(t) +
   \int_0^t \, 
 \int_{-\infty}^{+\infty}  \varphi(x-x',t-t')\, p(x',t')\, dt'\,dx',
\end{equation}
where $\delta(x)$ is Dirac's delta function and $\Psi(\tau)$ is the so-called survival function. $\Psi(\tau)$ is related to the marginal waiting-time probability density $\psi(\tau)$. The two marginal densities $\psi(\tau)$ and $\lambda(\xi)$ are:
\begin{eqnarray}
\label{marginal}
\psi(\tau) & = & \int_{-\infty}^{+\infty} \varphi(\xi, \tau) \, d \xi \nonumber \\
\lambda(\xi) & = & \int_{0}^{\infty} \varphi(\xi, \tau) \, d \tau,
\end{eqnarray}
and the survival function $\Psi(\tau)$ is:
\begin{equation}
\label{survival}
\Psi(\tau) = 1 - \int_{0}^{\tau} \psi (\tau') \, d \tau' = \int_{\tau}^{\infty} \psi (\tau') \, d \tau'.
\end{equation}
Both the two marginal densities and the survival function can be empirically derived from tick-by-tick financial data in a direct way.

The integral equation, eq. (\ref{masterequation}), can be solved in the Laplace-Fourier domain. The Laplace transform, $\widetilde{g}(s)$ of a (generalized) function $g(t)$ is defined as:
\begin{equation}
\label{laplacetransform}
\widetilde{g}(s) = 
\int_{0}^{+ \infty} dt \; 
\hbox{e}^{ -st} \, g(t)\,,
\end{equation}
whereas the Fourier transform of a (generalized) function $f(x)$ is defined as:
\begin{equation}
\label{fouriertransform}
\widehat {f}(\kappa) = 
\int_{- \infty}^{+ \infty} dx \, 
\hbox{e}^{i \kappa x} \, f(x)\,.
\end{equation}
A generalized function is a distribution (like Dirac's $\delta$) in the sense of S. L. Sobolev and L. Schwartz \cite{gelfand58}.

One gets:
\begin{equation}
\label{gensol}
\widetilde{\widehat p}(\kappa, s) = \widetilde \Psi(s)\, \frac{1}{1-\widetilde{\widehat \varphi}(\kappa, s)},
\end{equation}
or, in terms of the density $\psi(\tau)$:
\begin{equation}
\label{vargensol}
\widetilde{\widehat p}(\kappa, s) = \frac{1 -\widetilde \psi(s)}{s}\, \frac{1}{1-\widetilde{\widehat \varphi}(\kappa, s)},
\end{equation}
as, from eq. (\ref{survival}), one has:
\begin{equation}
\label{survivallt}
\Psi(s) = \frac{1 -\widetilde \psi(s)}{s}.
\end{equation} 
In order to obtain $p(x,t)$, it is then necessary to invert its Laplace-Fourier transform $\widetilde{\widehat p}(\kappa, s)$. As we shall see in the next subsection, for log-returns independent from waiting times, it is possible to derive a series solution to the integral equation (\ref{masterequation}).

\subsection{Limit Theorems: The Uncoupled Case}

In a recent paper, Gorenflo, Mainardi and the present author have discussed the case in which log-returns and waiting times are independent \cite{scalas04}. It is the so-called uncoupled case, when it is possible to write the joint probability density of log-returns and waiting times as the product of the two marginal densities:
\begin{equation}
\label{factorization}
\varphi(\xi, \tau) = \lambda(\xi) \psi (\tau)
\end{equation}
with the normalization conditions $\int d \xi \lambda (\xi) = 1$ and $\int d \tau \psi(\tau)=1$.

In this case the integral master equation for $p(x,t)$ becomes:
\begin{equation}
\label{uncoupledreal}
p(x,t) =  \delta (x)\, \Psi(t) +
   \int_0^t   \psi(t-t') \, \left[
 \int_{-\infty}^{+\infty}  \lambda(x-x')\, p(x',t')\, dx'\right]\,dt'
\end{equation}
This equation has a well known general explicit solution in terms of $P(n,t)$, the probability of $n$ jumps occurring up to time $t$, and of the $n$-fold convolution of the jump density, $\lambda_n (x)$: 
\begin{equation}
\label{jumpconv}
\lambda_n (x) = \int_{-\infty}^{+\infty} \ldots \int_{-\infty}^{+\infty} \,d \xi_{n-1} \ldots d \xi_1 \lambda(x - \xi_{n-1}) \ldots \lambda(\xi_1).
\end{equation}
Indeed, $P(n,t)$ is given by:
\begin{equation}
\label{Poisson1}
P(n,t) = \int_{0}^{t} \psi_n (t - \tau) \Psi(\tau) \, d \tau
\end{equation}
where $\psi_n (\tau)$ is the $n$-fold convolution of the waiting-time density: 
\begin{equation}
\label{timeconv}
\psi_n (\tau) = \int_{0}^{\tau} \ldots \int_{0}^{\tau_1}
\, d \tau_{n-1} \ldots d \tau_1 \psi(\tau-\tau_{n-1}) \ldots \psi(\tau_1).   
\end{equation}
The $n$-fold convolutions defined above are probability density functions for the sum of $n$ independent variables.

Using the Laplace-Fourier method and recalling the properties of Laplace-Fourier transforms of convolutions, one gets the following solution of the integral equation
\cite{Weiss BOOK94,scalas04,Mainardi FRACTAL04,Mainardi VIETNAM04}:
\begin{equation}
\label{series2}
p(x,t) = \sum_{n=0}^{\infty} P(n,t) \lambda_n (x)
\end{equation}
Eq. (\ref{series2}) can also be used as the starting point to derive eq. (\ref{uncoupledreal}) via the transforms of Fourier and Laplace, as it describes a jump process subordinated to a renewal process \cite{Clark 73,geman01}.

Let us now consider the following pseudodifferential equation, giving rise to anomalous relaxation and power-law tails in the waiting-time probability:
\begin{equation}
\label{anrelax}
\frac{d^\beta}{d \tau^\beta} \Psi(\tau) = - \Psi(\tau), \,\,\, \tau>0, \,\,\, 0 < \beta \leq 1; \,\,\, \Psi(0^+) =1,
\end{equation}
where the operator $d^\beta / dt^\beta$ is the Caputo fractional derivative, related to the Riemann--Liouville fractional derivative. For a sufficiently well-behaved function $f(t)$, the Caputo derivative is defined by the following equation, for $0<\beta<1$:
\begin{equation}
\label{Capder}
\frac{d^\beta}{dt^\beta} f(t) = \frac{1}{\Gamma(1-\beta)} \, \frac{d}{dt} \int_{0}^{t} \frac{f(\tau)}{(t-\tau)^\beta}\,d\tau - \frac{t^{-\beta}}{\Gamma(1-\beta)}f(0^{+}),
\end{equation}
and reduces to the ordinary first derivative for $\beta = 1$. The Laplace transform of the Caputo derivative of a function $f(t)$ is:
\begin{equation}
\label{Capderlt}
{\cal{L}} \left( \frac{d^\beta}{dt^\beta} f(t);\,s \right) = s^{\beta} \widetilde f (s) - s^{\beta - 1} f(0^+).
\end{equation}
If eq. (\ref{Capderlt}) is applied to the Cauchy problem of eq. (\ref{anrelax}), one gets:
\begin{equation}
\label{anrelaxlt}
\widetilde \Psi (s) = \frac{s^{\beta - 1}}{1+s^\beta}.
\end{equation}
Eq. (\ref{anrelaxlt}) can be inverted, giving the solution of
eq. (\ref{anrelax}) in terms of the Mittag-Leffler function of parameter $\beta$ \cite{mainardi96,gorenflo97}:
\begin{equation}
\label{ML}
\Psi(\tau) = E_{\beta} (-\tau^{\beta}),
\end{equation}
defined by the following power series in the complex plane:
\begin{equation}
\label{MLseries}
E_{\beta} (z) := \sum_{n=0}^{\infty} \frac{z^n}{\Gamma(\beta n +1)}.
\end{equation}
The Mittag-Leffler function is a possible model for a fat-tailed survival function.
For $\beta=1$, the Mittag-Leffler function coincides with the ordinary exponential
function. For small $\tau$, the Mittag-Leffler survival function coincides with the stretched exponential:
\begin{equation}
\label{stretched}
\Psi(\tau) = E_{\beta} (-\tau^{\beta}) \simeq 1 - \frac{\tau^\beta}{\Gamma(\beta +1)} \simeq \exp \{-\tau^\beta / \Gamma(\beta +1) \}, \; 0 \leq \tau \ll 1,
\end{equation}
whereas for large $\tau$, it has the asymptotic representation:
\begin{equation}
\label{powerlaw}
\Psi(\tau) \sim \frac{\sin (\beta \pi)}{\pi} \, \frac{\Gamma(\beta)}{\tau^\beta}, \; 0<\beta<1,
\; \tau \to \infty.
\end{equation}
Accordingly, for small $\tau$, the probability density function of waiting times $\psi(\tau) = - d \Psi(\tau) / d \tau$ behaves as:
\begin{equation}
\label{stretchedd}
\psi(\tau) = - \frac{d}{d \tau} E_{\beta} (-\tau^{\beta}) \simeq \frac{\tau^{-(1-\beta)}}{\Gamma(\beta)}, \; 0 \leq \tau \ll 1,
\end{equation}
and the asymptotic representation is:
\begin{equation}
\label{powerlawd}
\psi(\tau) \sim \frac{\sin (\beta \pi)}{\pi} \, \frac{\Gamma(\beta+1)}{\tau^{\beta+1}}, \; 0<\beta<1, \; \tau \to \infty.
\end{equation}

The Mittag-Leffler function is important as, without passage to the diffusion limit, it leads to a time-fractional master equation, just by insertion into the CTRW integral equation. This fact was discovered and made explicit for the first time in 1995 by Hilfer and Anton \cite{hilfer95b}. Therefore, this special type of waiting-time law (with its particular properties of being singular at zero, completely monotonic and long-tailed) may be best suited for approximate CTRW Monte Carlo simulations of fractional diffusion.

For processes with survival function given by the Mittag-Leffler function, the solution of the master equation can be explicitly written:
\begin{equation}
\label{exact}
p(x,t) = \sum_{n=0}^{\infty} \frac{t^{\beta n}}{n!} E_{\beta}^{(n)} (-t^\beta) \lambda_n(x),
\end{equation}
where:
$$E_{\beta}^{(n)}(z) := \frac{d^n}{dz^n} E_{\beta} (z).$$

The Fourier transform of eq. (\ref{exact}) is the characteristic function of 
$p(x,t)$ and is given by:
\begin{equation}
\label{characteristic}
\hat{p} (\kappa,t) = E_{\beta} [t^{\beta} (\hat{\lambda} (\kappa) - 1)].
\end{equation}
If log-returns and waiting times are scaled according to:
\begin{equation}
\label{scalingjumps}
x_n (h) = h \xi_1 + h \xi_2 + \ldots + h \xi_n,
\end{equation}
and:
\begin{equation}
\label{scalingtime}
t_n (r) = r \tau_1 + r \tau_2 + \ldots + r \tau_n,
\end{equation}
the scaled characteristic function becomes:
\begin{equation}
\label{scaledcharacteristic}
\hat{p}_{h,r} (\kappa,t) = E_{\beta} \left[ \frac{t^{\beta}}{r^{\beta}} 
(\hat{\lambda} (h \kappa) - 1) \right].
\end{equation}
Now, if we assume the following asymptotic behaviours for vanishing $h$ and $r$:
\begin{equation}
\label{limitstructure}
\hat{\lambda} (h \kappa) \sim 1 - h^{\alpha} |\kappa|^{\alpha}; \; \;
0 < \alpha \leq 2,
\end{equation}
and
\begin{equation}
\label{limitscaling}
\lim_{h,r \to 0} \frac{h^{\alpha}}{r^{\beta}} = 1,
\end{equation}
we get that:
\begin{equation}
\label{limitgreen}
\lim_{h,r \to 0} \hat{p}_{h,r} (\kappa, t) = \hat{u} (\kappa, t) = E_{\beta} [-t^{\beta} 
|\kappa|^{\alpha}].
\end{equation}
The Laplace transform of eq. (\ref{limitgreen}) is:
\begin{equation}
\label{GreenFCT}
\widetilde{\widehat u}(\kappa, s) = \frac{s^{\beta-1}}{|\kappa|^{\alpha}+s^{\beta}}.
\end{equation}
Therefore, the well-scaled limit of the CTRW characteristic function coincides with
the Green function of the following pseudodifferential {\it fractional} diffusion equation:
\begin{equation}
\label{FLfractional}
|\kappa|^{\alpha} \widetilde{\widehat u}(\kappa, s) + s^{\beta} \widetilde{\widehat u}(\kappa, s) = s^{\beta - 1},
\end{equation}
with $u(x,t)$ given by:
\begin{equation}
\label{greenfunction}
u(x,t) = \frac{1}{t^{\beta / \alpha}} W_{\alpha, \beta} \left( \frac{x}{t^{\beta/\alpha}} \right),
\end{equation}
where $W_{\alpha, \beta} (u)$ is given by:
\begin{equation}
\label{scalingfunction}
W_{\alpha, \beta} (u) = \frac{1}{2 \pi} \int_{-\infty}^{+\infty} \; d \kappa \; \hbox{e}^{-i \kappa u} E_\beta (-|\kappa|^\alpha),
\end{equation}
the inverse Fourier transform of a Mittag-Leffler function \cite{metzler00,mainardi01,zaslavsky02,scalas03,metzler04}.

For $\beta =1$ and $\alpha = 2$, the fractional diffusion equation reduces to the ordinary diffusion equation and the function $W_{2,1} (u)$ becomes the Gaussian probability density function evolving in time with a variance $\sigma^2 = 2t$. In the general case ($0 < \beta < 1$ and $0 < \alpha <2$), the function $W_{\alpha, \beta} (u)$ is still a probability density evolving in time, and it belongs to the class of Fox $H$-type functions that can be expressed in terms of a Mellin-Barnes integral as shown in details in ref. \cite{mainardi01}.

The scaling equation, eq. (\ref{limitscaling}), can be written in the following form:
\begin{equation}
\label{properscaling}
h \simeq r^{\beta/\alpha}.
\end{equation}
If $\beta = 1$ and $\alpha = 2$, one recognizes the scaling relation typical of Brownian motion (or the Wiener process). 

In the passage to the limit outlined above, $\widetilde{\widehat{p}}_{r,h} (\kappa, s)$ and $\widetilde{\widehat{u}} (\kappa, s)$ are asymptotically equivalent in the Laplace-Fourier domain. Then, the asymptotic equivalence in the space-time domain between the master equation and the fractional diffusion equation is due to the continuity theorem for sequences of characteristic functions, after the application of the analogous theorem for sequences of Laplace transforms \cite{feller71}. Therefore, there is convergence in law or weak convergence for the corresponding probability distributions and densities. Here, weak convergence means that the Laplace transform and/or Fourier transform (characteristic function) of the probability density function are pointwise convergent
(see ref. \cite{feller71}).

\subsection{Limit Theorems: The Coupled Case}

The diffusive limit in the coupled case is discussed by Meerschaert {\it et al.} \cite{meerschaert02}. The coupled case is relevant as, in general, log-returns and waiting times are not independent \cite{raberto02}. Based on the results summarized in \cite{scalas04} and discussed in \cite{gorenflo02,Gorenflo VIETNAM04}, it is possible to prove the following theorem for the coupled case: \\

\noindent {\bf Theorem} \\

\noindent Let $\varphi(\xi,\tau)$ be the (coupled) joint probability density of a CTRW. If, under the scaling $\xi \to h \xi$ and $\tau \to r \tau$, the Fourier-Laplace transform of $\varphi(\xi, \tau)$ behaves as follows:
\begin{equation}
\label{scaling1}
\widetilde{\widehat \varphi}_{h,r}(\kappa, s) = \widetilde{\widehat \varphi}(h \kappa, rs)  
\end{equation}
and if, for $h \to 0$ and $r \to 0$, the asymptotic relation holds:
\begin{equation}
\label{scaling2}
\widetilde{\widehat \varphi}_{h,r}(\kappa, s) = \widetilde{\widehat \varphi}(h \kappa, rs) \sim 1 - \mu |h \kappa|^{\alpha} - \nu (rs)^{\beta},
\end{equation}
with $0< \alpha \leq 2$ and $0 < \beta \leq 1$.
Then, under the scaling relation $\mu h^{\alpha} = \nu r^{\beta}$, the solution of the (scaled) coupled CTRW master (integral) equation, eq. (\ref{masterequation}), $p_{h,r} (x,t)$, weakly converges to the Green function of the fractional diffusion equation, $u(x,t)$, for $h \to 0$ and $r \to 0$. \\

\noindent {\it Proof} \\

\noindent The Fourier-Laplace transform of the scaled conditional probability density 
$p_{h,r} (x,t)$ is given by:
\begin{equation}
\widetilde{\widehat p}_{h,r} (\kappa, s) = \frac{1 - \widetilde{\psi}(rs)}{s} \frac{1}{1-\widetilde{\widehat \varphi}(h \kappa, r s)}.
\label{scaledlaplace}
\end{equation}
Replacing eq. (\ref{scaling2}) in eq. (\ref{scaledlaplace}) and observing that 
$\widetilde{\psi}(s) = \widetilde{\widehat \varphi}(0,s)$, one asymptotically gets
for small $h$ and $r$:
\begin{equation}
\label{asympt1}
\widetilde{\widehat p}_{h,r} (\kappa, s) \sim  \frac{\nu r^{\beta} s^{\beta -1}}{\nu r^{\beta} s^{\beta}+\mu h^{\alpha} |\kappa|^{\alpha}},
\end{equation}
which for vanishing $h$ and $r$, under the hypotheses of the theorem, converges to:
\begin{equation}
\label{asympt2}
\widetilde{\widehat p}_{0,0} (\kappa, s) = \widetilde{\widehat u}(\kappa, s) = \frac{s^{\beta -1}}{ s^{\beta}+ |\kappa|^{\alpha}},
\end{equation}
where $\widetilde{\widehat u}(\kappa, s)$ is the Fourier-Laplace transform of the Green
function of the fractional diffusion equation (see eq. (\ref{FLfractional})). The asymptotic equivalence in the space-time domain, between $p_{0,0} (x,t)$ and $u(x,t)$, the inverse Fourier-Laplace transform of $\widetilde{\widehat u}(\kappa, s)$, is again ensured by the continuity theorem for sequences of characteristic functions, after the application of the analogous theorem for sequences of Laplace transforms \cite{feller71}. There is convergence in law or weak convergence for the corresponding probability distributions and densities.

An important consequence of the above theorem is the following corollary showing that
in the case of marginal densities with finite first moment of waiting times and finite second moment of log-returns, the limiting density $u(x,t)$ is the solution of the ordinary diffusion equation (and thus the limiting process is the Wiener process). The corollary can be used to justify the popular Geometric Brownian Motion model of stock prices, here with expected return set to zero. Again, in order to derive this result, no reference is necessary to the Efficient Market Hypothesis \cite{fama70,fama91}. \\

\noindent {\bf Corollary} \\

\noindent If the Fourier-Laplace transform of $\varphi(\xi,\tau)$ is regular for $\kappa = 0$ and $s=0$, and, moreover, the marginal waiting-time density, $\psi(\tau)$, has finite first moment $\tau_0$ and the marginal jump density, $\lambda(\xi)$, is symmetric with finite second moment $\sigma^2$, then the limiting solution of the master (integral) equation for the coupled CTRW is the Green function of the ordinary diffusion equation. \\

\noindent{\it Proof} \\

\noindent Due to the hypothesis of regularity in the origin and to the properties of Fourier and Laplace transforms, we have that:
\begin{eqnarray}
\label{LFdiffusion1}
& &\widetilde{\widehat \varphi}_{h,r}(\kappa, s) = \widetilde{\widehat \varphi} (h \kappa, r s) \sim  \widetilde{\widehat \varphi} (0,0) + \nonumber  \\
& &+ \frac{1}{2} \left( \frac{\partial^{2} \widetilde{\widehat \varphi}}{\partial \kappa^{2}} \right)_{(0,0)} h^2 \kappa^2 + \left( \frac{\partial \widetilde{\widehat \varphi}}{\partial s} \right)_{(0,0)} rs = \nonumber \\
 & & = 1 - \frac{\sigma^{2}}{2} h^2 \kappa^2 - \tau_{0} r s,
\end{eqnarray}
and as a consequence of the theorem, under the scaling $h^2 \sigma^2 / 2 = \tau_0 r$,
one gets, for vanishing $h$ and $r$:
\begin{equation}
\label{LFdiffusion2}
\widetilde{\widehat p}_{0,0} (k,s) = \widetilde{\widehat u} (k,s) = \frac{1}{s+k^2},
\end{equation}
corresponding to the Green function (\ref{FLfractional}) for $\alpha = 2$ and $\beta =1$, that is the solution of the Cauchy problem for the ordinary diffusion equation. 

\section{Summary and Outlook}

In this paper, a discussion of continuous-time random walks (CTRWs) has been presented as phenomenological models of tick-by-tick market data. Continuous-time random walks are rather general and they include compound Poisson processes as particular instances. Well-scaled limit theorems have been presented for a rather general class of CTRWs.

It is the hope of this author that this paper will stimulate further research on high-frequency econometrics based on the concepts outlined above. There are several possible developments.

First of all, one can abandon the hypothesis of i.i.d. log-returns and waiting times and consider various forms of dependence. In this case, it is no longer possible to exploit the nice properties of Laplace and Fourier transforms of convolutions, but, still, Monte Carlo simulations can provide hints on the behaviour of these processes in the diffusive limit.

A second possible extension is to include volumes as a third stochastic variable. This extension is straightforward, starting from a three-valued joint probability density.

A third desirable extension is to consider a multivariate rather than univariate model that includes correlations between time series.

The present author is currently involved in these extensions and is eager to know progress in any direction by other independent research groups. He can be contacted at {\tt scalas@unipmn.it}

\section{Acknowledgements}

This work was supported by the Italian M.I.U.R. F.I.S.T. Project ``High frequency dynamics in financial markets''. 
The author wishes to acknowledge stimulating discussion with P. Buchen, S. Focardi, R. Gorenflo, T. Kaizoji, H. Luckock, F. Mainardi, M. M. Meerschaert and M. Raberto.

\printindex
\end{document}